\documentclass[aps,prl,reprint,groupedaddress,showpacs]{revtex4-1}

\usepackage{graphicx}
\usepackage{dcolumn}
\usepackage{bm}
\usepackage{color}

\begin{document}



\title{Intermittent Flow In Yield-Stress Fluids Slows Down
Chaotic Mixing}


\author{D. M. Wendell}
\author{F. Pigeonneau}
\author{E. Gouillart}
\author{P. Jop}
\affiliation{Surface du Verre et Interfaces, UMR 125 CNRS/Saint-Gobain, 39, quai Lucien Lefranc, F-93303 Aubervilliers, Cedex, France}


\date{\today}

\begin{abstract}

In this article, we present experimental results of chaotic mixing of
Newtonian fluids and yield stress fluids using rod-stirring protocol with
rotating vessel.  We show how the mixing of yield stress fluids by chaotic advection
is reduced compared to the mixing of Newtonian fluids and explain
our results bringing to light the relevant mechanisms: the
presence of fluid that only flows intermittently, a phenomenon enhanced 
by the yield stress, and the importance of the peripheral region.
This finding is confirmed via numerical
simulations. Anomalously slow mixing is observed when the synchronization
of different stirring elements leads to the repetition of slow stretching
for the same fluid particles.

\end{abstract}


\maketitle

\begin{figure*}
\includegraphics[width = 6.75in]{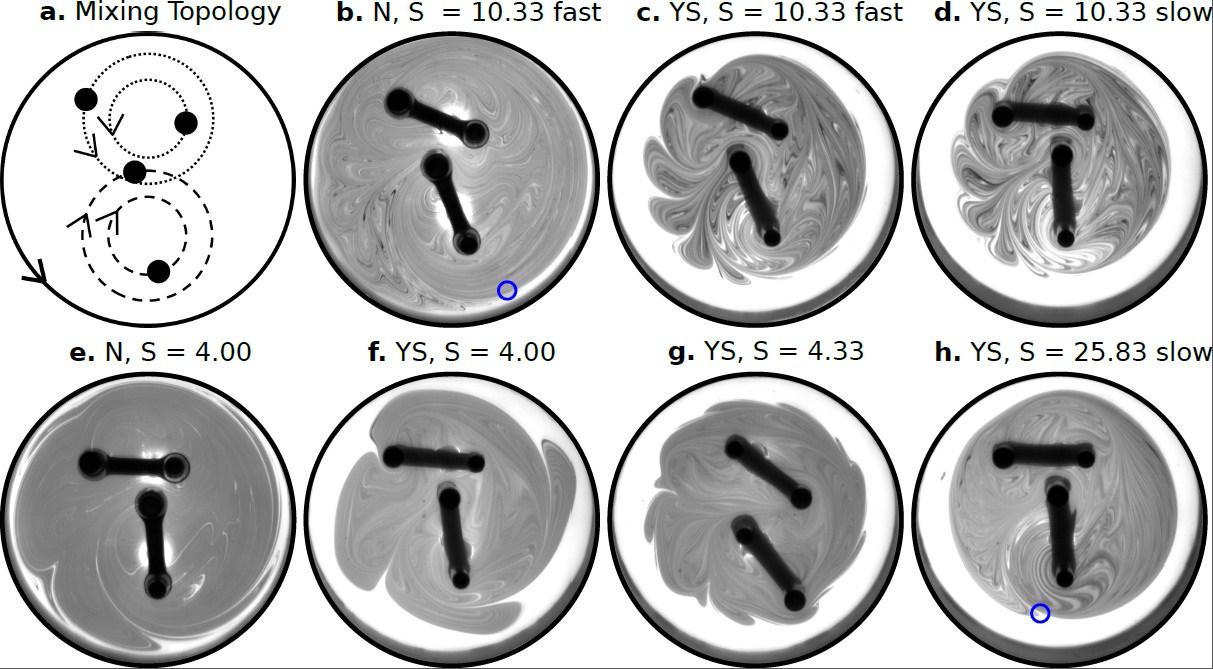}%

\caption{(Color online) The mixing protocol and selected results. All times, $t$, are
normalized by the period of the stirring rods. The fluid composition is
indicated by N for Newtonian fluid and YS for yield stress fluid. [a] The
stirring protocol.  Note the four rotating stirring rods and the
direction of rotation of the vessel. [b] Newtonian fluid, $S$=10.33, fast
velocity, $t$=20. [c] 0.1~w.t.\% YS fluid, $S$=10.33, fast velocity,
$t$=20. [d] 0.1~w.t.\% YS fluid, $S$=10.33, slow velocity, $t$=20. [e]
Newtonian fluid, $S$=4.00, $t$=50. [f] 0.1~w.t.\% YS fluid, $S$=4.00,
$t$=50. [g] 0.1~w.t.\% YS fluid, $S$=4.33, $t$=50. [h] 0.1~w.t.\% YS
fluid, $S$=25.83, $t$=40. The tiny blue circles represent fixed (1-periodic)
points. \label{pics}}

\end{figure*}

Many mixing situations involve fluids with non-Newtonian properties:
mixing of building materials such as concrete or mortar are based on
fluids that have shear-thinning rheological properties.  Lack of correct
mixing can waste time and money, or lead to products with
defects~\cite{arratia06}. When fluids are stirred and mixed together at
low Reynolds number, the fluid particles should undergo chaotic
trajectories to be well mixed by the so-called \emph{chaotic
advection}~\cite{aref84} resulting from the flow. Previous
work to characterize chaotic mixing in many different geometries has
primarily focused on Newtonian fluids~\cite{ottino90}. First studies
into non-Newtonian chaotic advection often utilize idealized mixing
geometries such as cavity flows or journal bearing flows for numerical
studies~\cite{anderson00, niederkorn94, fan01}.

In this article, we present results of mixing of Newtonian fluids and
yield-stress fluids.  We implement a mixing protocol inspired by the
figure-eight stirring with rotating vessel described in
\cite{gouillart10}. We observe the effects of the yield stress in the
characteristic shape of the mixing area and in the mixing rate when
varying the absolute velocity of the system. We show that the ratio of
rotational velocity between the stirring rods and the wall of the vessel
significantly influences the long-time mixing behavior of this mixing
protocol. Resonances at integer values of this ratio are shown to lead to
slow mixing. Numerical simulations are utilized to calculate the average
area below the yield stress. We find that areas of slow or intermittent
stretching due to the rheological properties of the fluid slow down
mixing of the whole system. 

We begin with a description of the apparatus.
All experiments investigate 2D flows with four periodically-driven
cylindrical stirring rods rotating with constant angular velocity in an
eggbeater configuration (Fig.~\ref{pics}a), designed to promote
chaotic advection, inside of a cylindrical mixing
vessel with a diameter of 14 cm.  The mixing vessel can also be rotated
at constant angular velocity. The importance of the vessel rotation has
been discussed in \cite{gouillart10, gouillart07, thiffeault11}. The
experimental parameters can be found in Table~\ref{mixtable}.  The
experiments are characterized by the stirring ratio, $S$, the ratio between
the period of rotation of the vessel and the period of rotation of the
stirring rods.  Three fluids were used in the experiments: cane sugar syrup, a
Newtonian fluid with a kinematic viscosity of
$5$~x~$10^{-4}$~m$^2$s$^{-1}$, and 0.1~w.t.\% and 0.2~w.t\% Carbopol
polymer solutions from Lubrizol, which are yield stress (YS) fluids
characterized by the Herschel-Bulkley law \cite{HBlaw}.  A spot of
low-diffusivity dye (Colorex trichromatic black, from Pebeo) is injected
in the middle depth of the vessel to minimize surface effects. We
follow the evolution of the dye concentration field during the mixing
process by taking photographs through the transparent bottom of the
mixing vessel once per period of the rods. In addition, numerical simulations are
realized with a finite-element
method. The constitutive equation of Herschel-Bulkley law is regularized
according the model of Deglo de Besses et al. \cite{DEGLOETAL2003}. The
area where the fluid is in solid rotation is extracted along one period
of rotation of the rods.

\begin{table}
\caption{Mixing Rate Experimental Parameters. The stirring ratio, $S$, is
the mixing vessel rotation period divided by the stirring rod rotation
period.\label{mixtable}}
\begin{ruledtabular}
\begin{tabular}{ccc}
\bf{$S$ Ratio} & \bf{Vessel Period [sec]} & \bf{Rod Period [sec]}\\
4.00 & 24 & 6 \\
4.33 & 26 & 6 \\
10.33 slow & 62 & 6 \\
10.33 fast & 31 & 3 \\
25.83 slow & 155 & 6 \\
\end{tabular}
\end{ruledtabular}
\end{table}

The topology of the mixing region has several interesting features.
Representative photographs are shown in Fig.~\ref{pics}b-h. The fluid
domain is separated into a large chaotic region delineated by the
long-term dye pattern, and a non-chaotic dye-free region where particles
are entrained on closed trajectories by the rotating
wall~\cite{gouillart10}. For the yield-stress fluid, the width of the
non-chaotic region is larger than for the Newtonian fluid because of a
zone of intermittent solid rotation close to the rotating vessel. Its
boundary corresponds roughly to the largest extent of the rods'
trajectories. For a slow velocity of the vessel, the mixing pattern
displays a single injection point (blue circle in Fig.~\ref{pics}b and
\ref{pics}h). The stroboscopic position of this point is fixed at every period of
the rods, and stems from the competition between the moving wall and
the lower pair of rods. When the speed of the wall increases, this fixed
(periodic) point is shifted within the chaotic region. Then, fluid
particles inside the chaotic region but close to its boundary circle
along the rotating wall for a few periods, before escaping to the center
of the mixing region. In this case, the border of the chaotic region
shows several \emph{lobes}, that are \emph{almost-cyclic
sets}~\cite{stremler11} delineated by dye-free
cusps~\cite{thiffeault08} (Fig.~\ref{pics}c-g). For the yield-stress
fluid, fluid inside the lobes is only intermittently sheared, due to the
proximity of the wall. The number of lobes is approximately equal
to the stirring ratio. The transition from one fixed point to travelling
lobes occurs earlier when increasing the wall speed, {\it i.e.} at larger $S$, for the yield-stress fluid than for the
Newtonian fluid (Fig.~\ref{pics}b vs. \ref{pics}c). This is
because the angular velocity stays almost constant in the quasi solid-rotation
zone of the yield-stress fluid, but this zone is sheared for the
Newtonian fluid and the angular velocity profile decays much faster with
the distance to the wall. 

\begin{figure}
\includegraphics[width = 3.5in]{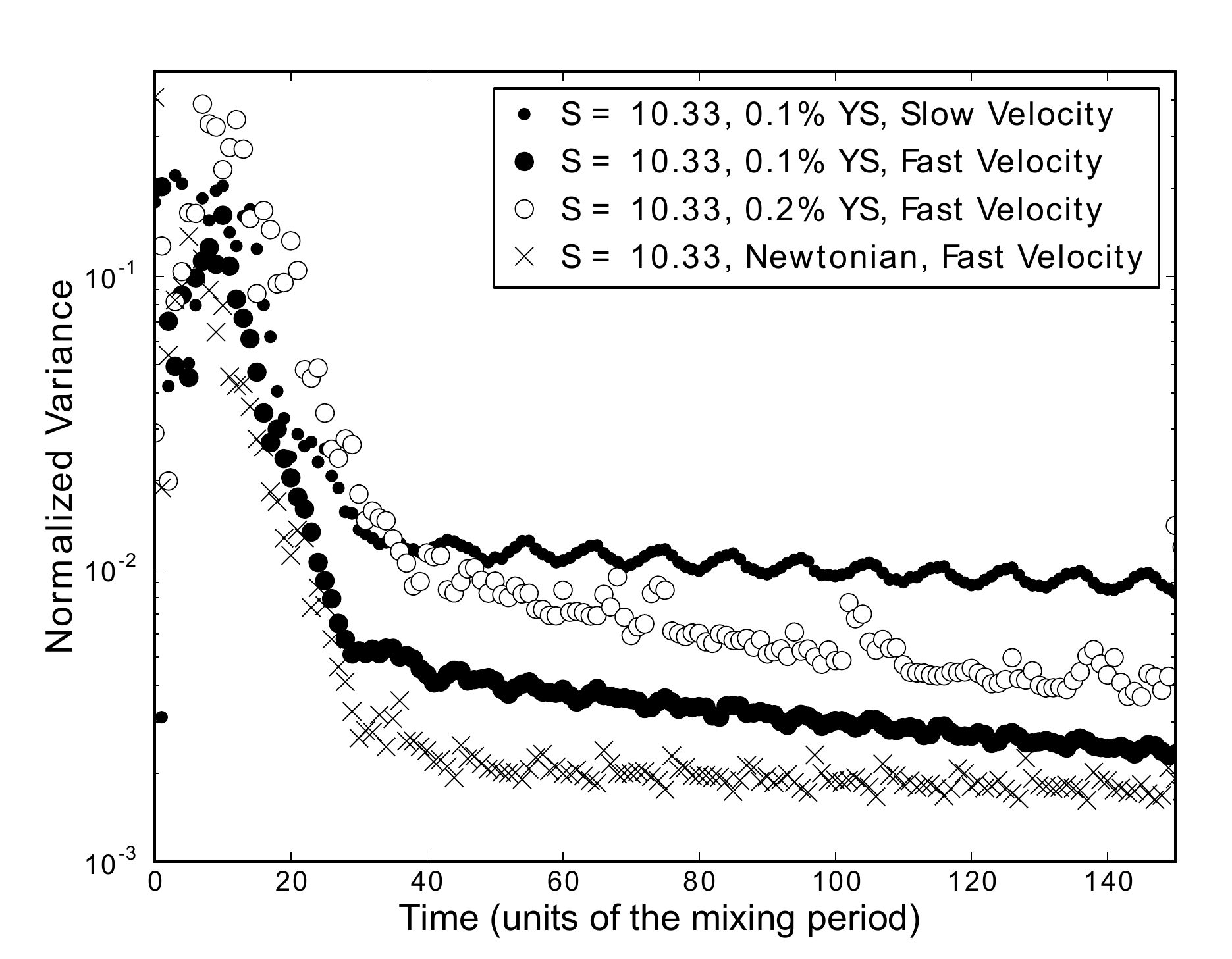}%
\caption{Variance for different fluids mixed with a stirring ratio of
$S$=10.33.  The same stirring ratio can give different mixing
rates based on the rheology of the fluid and the stirring velocity.
Representative mixing images are seen in
Fig.~\ref{pics}b-d.\label{S10.3}}
\end{figure}

The images are processed to calculate the variance of the dye
concentration inside the mixing region, and to measure the mixing
region's area. In the variance measurements, we observe two distinct
regimes of mixing (see Fig.~\ref{S10.3}). Initially, a rapid
exponential mixing occurs where the dye is quickly stretched and folded
until it almost fills the entire mixing region. Then a slower exponential
mixing regime occurs where a minority of dye filaments that are not yet
stretched to the width of the diffusion length are progressively refined
through additional stretching and folding \cite{graphsendnote}. From
these two regimes, we can define two indices of the quality of mixing:
(1) the level of mixing accomplished at the end of the first, rapid
mixing regime and (2) the slope in the slower regime, i.e., the rate of
mixing at long times. In the slower regime, the mixing pattern has
converged to the slowest decaying advection-diffusion eigenmode
\cite{sukhatme02, gouillart10}, and the decay rate is the corresponding
eigenvalue. 

The effect of the yield stress is observed in experiments with different
fluid rheologies where the $S$ ratio is constant, but the absolute velocity
changes. Fig.~\ref{S10.3} shows results for several experiments each
with an $S$ ratio of 10.33. For the same yield stress fluid, increasing
the velocity increases the stress in the fluid, which leads to an
increase in the amount of fluid flowing during one period.
Experimentally, doubling the velocity led to increased mixing during both
the initial fast regime and secondary slower mixing regime. The origin of
this difference is apparent on the dye patterns of Fig.~\ref{pics}(c)
and \ref{pics}(d): the contrast between the core of the mixing region and the lower
peripheral part of the pattern is higher for the slow velocity. This
reveals poor transport between the two regions, due to the even more
intermittent stretching in the case of the slow velocity. Similarly, using
the same mixing velocity but a higher yield stress value fluid shows that
the higher yield stress fluid mixes slower.  The Newtonian fluid mixes
better than any of the yield stress fluids.

\begin{figure}
\includegraphics[width = 3.5in]{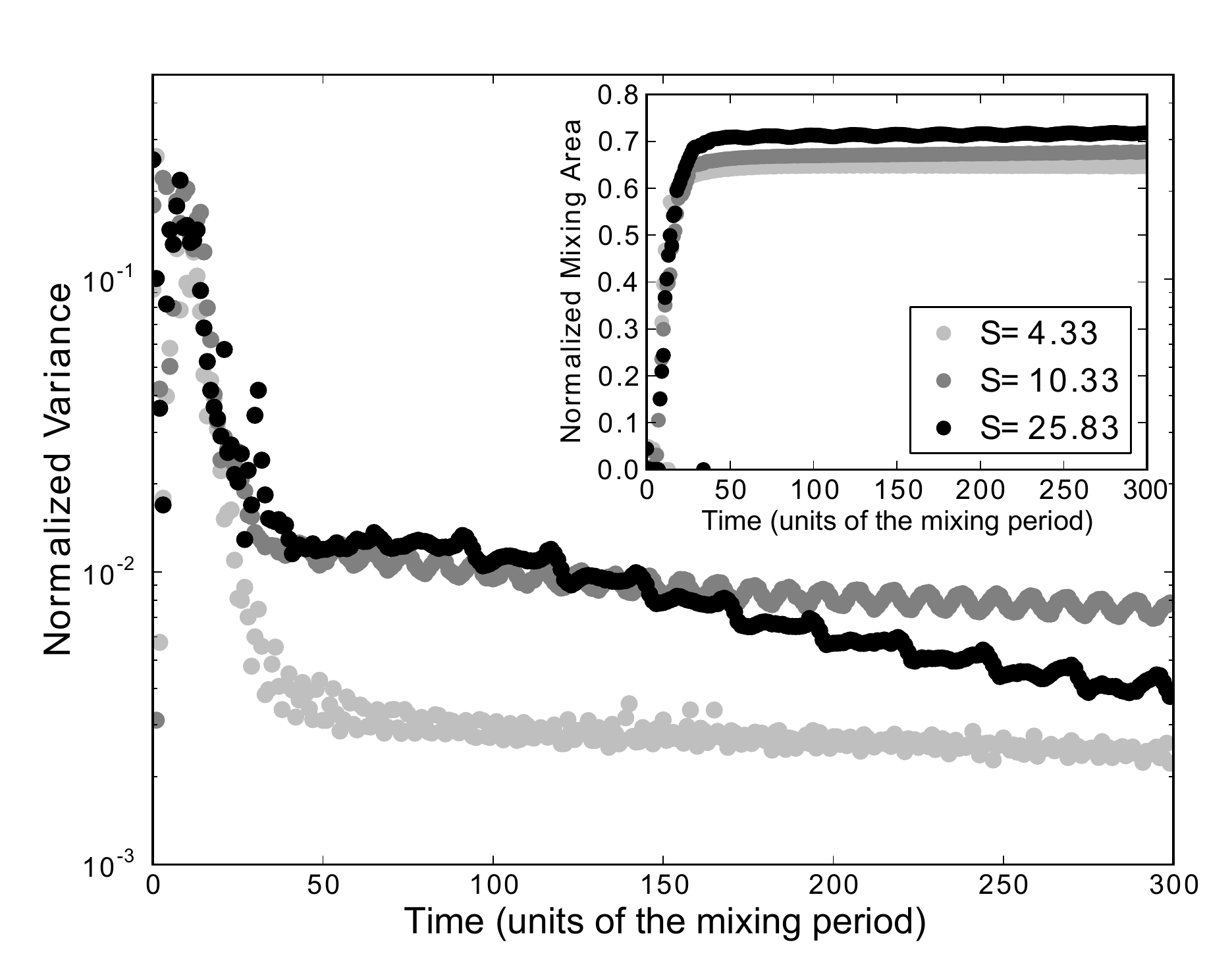}%

\caption{Mixing rates of 0.1\% YS fluid for three different vessel wall
speeds, all with a stirring rod period of 6 sec. After the first mixing
regime, the smaller $S$ ratio is the best mixed, but the rate of mixing
is higher with larger $S$ ratios at long time. Inset: mixing area as a
fraction of total vessel area for each of the $S$ ratios. Typical
patterns can be seen in Fig.~\ref{pics}d,g,h. \label{diff-ratios}}

\end{figure}

We now investigate the effect of wall speed on the mixing dynamics. It was
shown in \cite{gouillart10} that the decay rate of the variance in a
Newtonian fluid and in the single injection point regime increases with
increasing wall speed.  However, we observe a different situation in
experiments with 0.1\% YS fluid, where lobes are observed at larger
S-ratio. Fig.~\ref{diff-ratios} shows the variance of experiments with
three different wall speeds. Although at the end of the rapid mixing
regime the lowest $S$ ratio has mixed the best, it is clear during the
second slower mixing regime that the long-time rate of mixing increases
with increasing $S$ ratio. Hence, a high long-time decay rate seems to
correlate with a high level of variance when the second phase begins.
This can be traced back to the size of the slow-stretching peripheral
region: tiny lobes are observed for $S$=4.33, larger lobes for $S$=10.33, and
a large cusp, for $S$=25.83, displays important concentration fluctuations on both sides
of the injection point~\cite{gouillart10} (see Fig. 1g,d,h). This peripheral region
represents the region of slowest stretching -- because of intermittent
flow of the yield-stress fluid -- of the chaotic region, and the
long-time decay is controlled by the transport of unmixed fluid from the
peripheral region to the center. The larger the peripheral region, the
faster the transport rate, but the higher the quantity of unmixed fluid
transported to the center. The inset in Fig.~\ref{diff-ratios} confirms that
the mixing area increases with increasing $S$ ratio. 

\begin{figure}
\includegraphics[width = 3.5in]{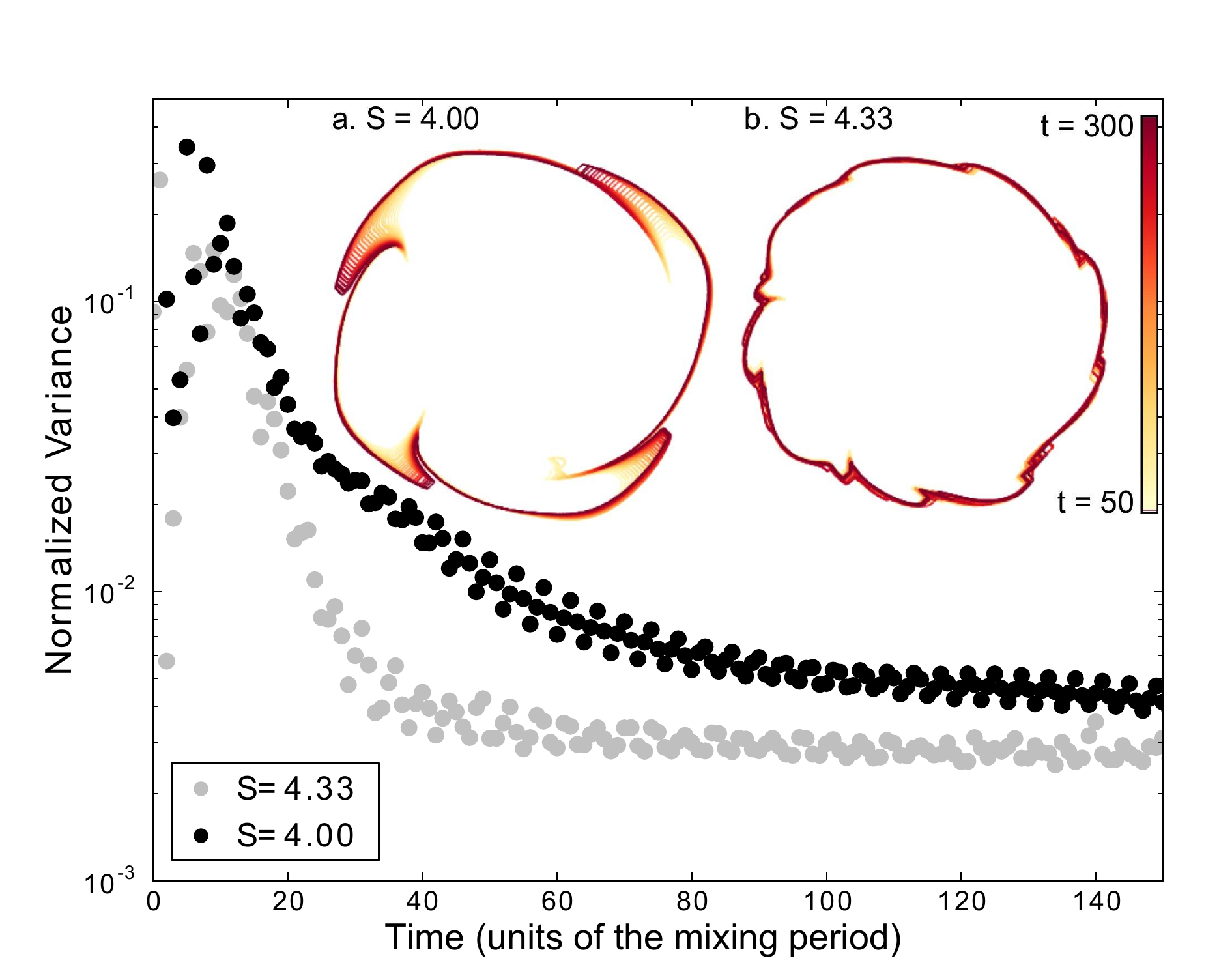}%
\caption{(Color online) Mixing rates of 0.1\% YS fluid for two different vessel wall
speeds. Note the slow mixing for $S=4.00$ as compared to the
$S=4.33$ case. Representative mixing images can be seen in
Fig.~\ref{pics}f,g. Inset: time evolution of the edge of the mixing area,
every 4 rotation
periods of the stirring rods, periods ($t$) 50 through 300.
\label{S4andS433}}
\end{figure}

Important changes in mixing quality can also result from a small
modification of the $S$ ratio, as shown in Fig.~\ref{S4andS433}. If we
increase further the wall speed from an $S$ ratio of 4.33 to an $S$ ratio of 4.00, we
observe an intermediate mixing regime which decreases the efficacy of
mixing in Fig.~\ref{S4andS433}.  This additional regime at $S$=4 comes from
a resonance phenomenon: at the boundary of the mixing region, some fluid
particles are entrained in solid rotation with the vessel, and are
synchronized with the rod's motion because of the integer stirring ratio.
As a result, the tip of the four lobes in Fig.~\ref{pics}f dodges
repeatedly the shearing action of the rods, resulting in regions where
stretching is anomalously slow. This is confirmed by the time-evolution
of the mixing area boundary, delineated in Fig.~\ref{S4andS433} (inset):
whereas the final shape is reached within 50
periods for $S$=4.33, in the $S$=4.00 case the final mixing
area is not achieved until 150 periods, due to
the slow development of lobes. In
contrast, for the non-integer ration $S$=4.33, peripheral fluid particles
are not synchronized with the rods and experience a succession of low and
high stretchings that self-average randomly~\cite{Villermaux2003} over
the stirring periods. Sharp changes of the mixing rate at integer
stirring ratios have also been observed in numerical simulations of
temperature transport~\cite{Lester2009}: for scalar transport, we propose
here that this resonance is due to the repetition of slow stretching
events.

\begin{figure}
\centerline{\includegraphics[width=.99\columnwidth]{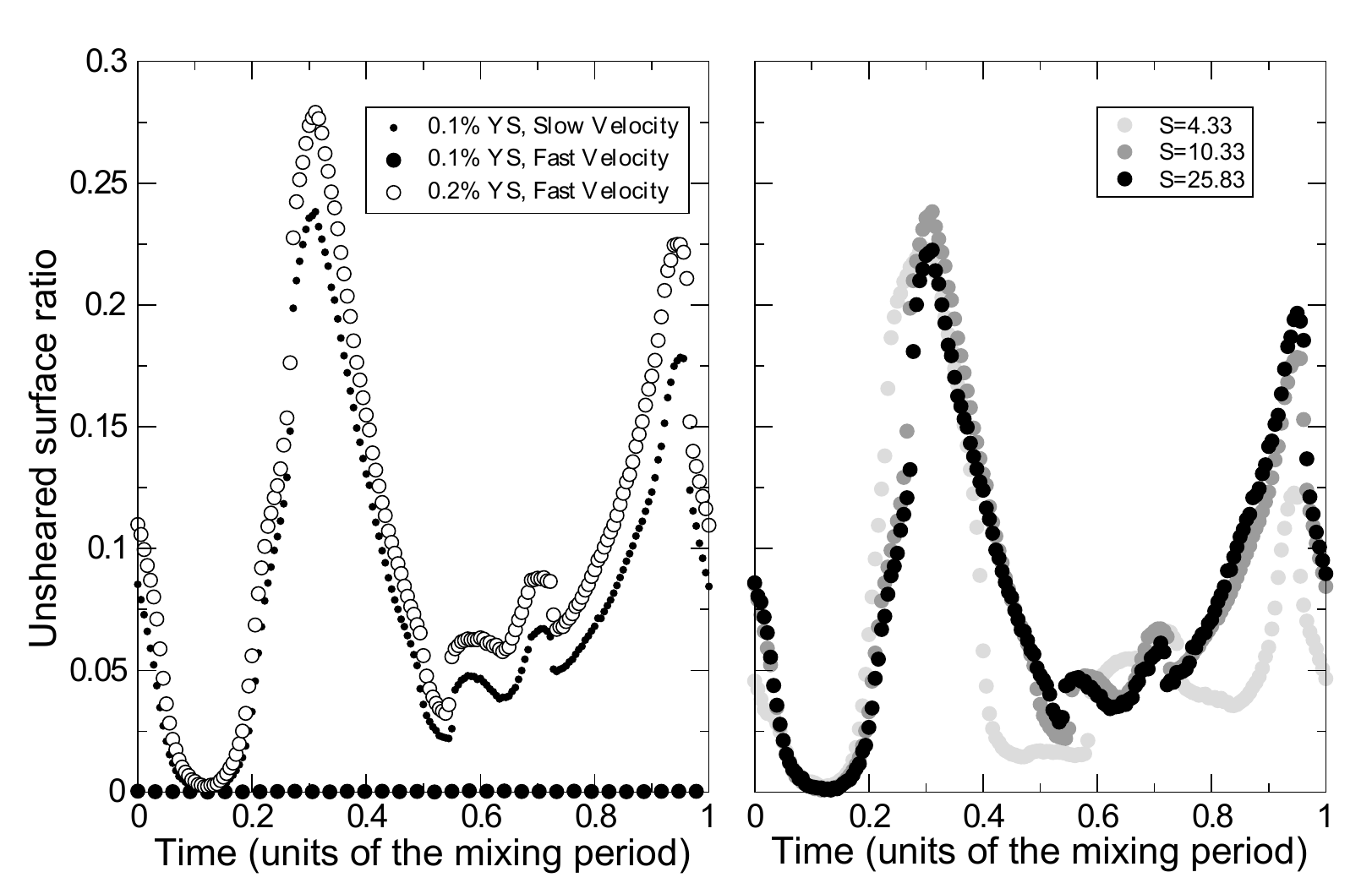}}
\caption{Unsheared surface fraction (the local shear stress is below the yield
stress) of a disk of radius 6.3 cm obtained in numerical simulations vs. time, during one
stirring period. (a) Same conditions as Fig.~\ref{S10.3} (different
rheologies). (b) Same conditions as Fig.~\ref{diff-ratios} (different
wall speeds).\label{fig:simus}}
\end{figure}

We have investigated in numerical simulations how the fraction of the
mixing region inside a circle of radius 6.3 cm and below the yield stress (shown in Fig.~\ref{fig:simus})
correlates with the decay of the variance observed in experiments. The
value of this non-flowing fraction fluctuates strongly during one
stirring period, since the distance between the rods and the wall varies.
For the fast velocity of the $0.1\%$ YS fluid, all the fluid is sheared at all
times in this region. When averaged over one period, the
non-flowing fraction ranks protocols in the same order as the level of
the concentration variance at the end of the fast mixing regime.
Indeed, the first fast mixing regime is well described by the global transport within
the mixing region. However, this sorting is no longer valid for the long-time decay rate (Fig.~\ref{fig:simus}
and Fig.~\ref{S10.3}).
This surprising result arises from slight Lagrangian correlations of low-stretching regions
governing the long-time decay. In these chaotic flows, such correlations are hard
to predict but in the resonant case (e.g. $S=4.00$), where correlations are enhanced. Therefore a
simple link does not exist between rheological properties and long time efficiency of the mixing protocol.

The rheological properties play an important role in the efficacy of
chaotic mixing.  We have shown that the decreasing efficiency using
chaotic advection to mix yield stress fluids compared to Newtonian
fluids is due to a major mechanism: the Lagrangian correlation of
intermittent flow of fluid when it is below the yield stress. Such
fluid particles are found in particular at
the periphery of the chaotic region, where the rotating wall entrains
particles in solid rotation, except when they are sheared transiently by
the passing rods. At long time, the exponential mixing rate is controlled
by the transport of fluid between the peripheral region and the center.
Resonances at integer stirring ratios allow some fluid particles of the
chaotic region to experience repeatedly a very low stretching value,
resulting in anomalously slow transport and mixing. Hence, integer
ratios should be avoided for practical applications. In future work, we
plan to compare numerically the Lagrangian statistics of stretching
for the different protocols.

\begin{acknowledgments}
The authors acknowledge support from the ANR (project Rheomel
ANR-11-JS09-015) and helpful discussions with Ph. Coussot.
\end{acknowledgments}


\begin{thebibliography}{17}
\bibitem [{HBl()}]{HBlaw}%
  \BibitemOpen
  \href@noop {} {}\bibinfo {howpublished} {Hershel-Bulkley fluid rheology
  presents a yield stress and follows the stress equation $\tau = \tau_0 +
  \kappa \dot{\gamma}^{n}$ where $\tau$ is the stress, $\tau_0$ is the yield
  stress, $\kappa$ is the consistency value, $\dot{\gamma}$ is the shear rate
  and $n$ is the flow index. The measured values were $\tau_0$ = 3\,Pa,
  $\kappa$ = 4\,Pa.s$^n$, and $n$ = 0.45 for the 0.1\,w.t.\% solution of
  Carbopol, and $\tau_0$ = 18\,Pa, $\kappa$ = 15\,Pa.s$^n$, and $n$ = 0.4 for
  the 0.2\,w.t.\% solution.}\BibitemShut {Stop}%
\bibitem [{gra()}]{graphsendnote}%
  \BibitemOpen
  \href@noop {} {}\bibinfo {howpublished} {A periodic variation occurs in each
  experiment. This periodicity comes from the period of rotation of the mixing
  vessel and a slight misalignment between the centerline of the mixing vessel
  and its rotating axis.}\BibitemShut {Stop}%
\end{thebibliography}

\providecommand{\noopsort}[1]{}\providecommand{\singleletter}[1]{#1}%

\end{document}